\documentclass[twocolumn,showpacs,preprintnumbers,amsmath,amssymb]{revtex4-2} \usepackage{dcolumn}

\usepackage{graphicx} \usepackage{bm}

\newcommand{\bma}{\mathbf a}
\newcommand{\bmb}{\mathbf b}
\newcommand{\bmdelta}{\bm \delta}
\newcommand{\bmK}{\mathbf K}
\newcommand{\bmr}{\mathbf r}
\newcommand{\bmR}{\mathbf R}

\newcommand{\bmx}{\mathbf x}
\newcommand{\bmy}{\mathbf y}
\newcommand{\bmZero}{\mathbf 0}

\newcommand{\calM}{{\cal M}}
\newcommand{\eK}{\epsilon_K}
\newcommand{\hg}{\hat{g}}

\newcommand{\hG}{\hat{G}}
\newcommand{\hH}{\hat{H}}
\newcommand{\hU}{\hat{U}}
\newcommand{\tm}{\text{-}}
\newcommand{\tgc}{\tilde{g_c}}

\begin{document}

\title{Green's function and LDOS for non-relativistic electron pair}
\date{\today}
\author{Tomasz M. Rusin} \email{email: tmr@vp.pl}
\affiliation{Institute of Physics, Polish Academy of Sciences, Al. Lotnik\'ow 32/46, 02-688 Warsaw, Poland}

\begin{abstract}

The Coulomb Green's function (GF) for non-relativistic charged particle in field of attractive Coulomb force
is extended to describe the interaction of two non-relativistic electrons
through repulsive Coulomb forces. Closed-form expressions for the GF,
in the absence of electron spins, are derived as one-dimensional integrals.
The results are then generalized to include electron spins and account for the Pauli exclusion principle.
This leads to a final GF composed of two components,
one even and the other odd with respect to exchange particles,
with closed-form expressions represented as one-dimensional integrals.
The Dyson equations for spin-independent potentials is presented.
The local density of states (LDOS) is calculated, which is a combination of contributions
from both even and odd GFs. This calculation reveals the dependence of LDOS
on inter-electron distance and energy. Separate analysis of the impact of 
the Pauli exclusion principle is provided.
An examination of the pseudo-LDOS, arising from the two-body contribution to the Green's
function, is undertaken. Complete suppression of the LDOS at~$r=0$ is ensured by this term,
which exhibits a restricted spatial extent. The reasons for the emergence of this pseudo-LDOS are elucidated.
\end{abstract}

\maketitle

\section{Introduction}

In 1963, Hostler and Pratt introduced a closed-form solution for the Green's
function (GF) of a non-relativistic
particle in the presence of an attractive Coulomb potential~\cite{Hostler1963}.
This GF is expressed in terms of Whittaker functions with complex arguments.
The derivation of this result was the culmination of extensive efforts,
which began with Meixner's work in 1933~\cite{Meixner1933}
and involved multiple approaches~\cite{Mapleton1961,Mano1963}.
Hostler subsequently re-derived the Coulomb GF using
various methods, yielding several equivalent
expressions~\cite{Hostler1964,Hostler1967,Hostler1970}. Hameka pursued a different approach and obtained the Coulomb GF
in an alternative form~\cite{Hameka1967,Hameka1968}.
Schwinger calculated the Coulomb GF in momentum space~\cite{Schwinger1964},
while Blinder derived it in parabolic coordinates~\cite{Blinder1981},
extending Hostler's approach to cover repulsive Coulomb potentials.
Swainson and Drake further extended Hostler's results to the relativistic case~\cite{Swainson1991}.
For a comprehensive review of papers related to the Coulomb GF and its applications
in multi-photon process calculations, Maquety {\it et al.} provide an insightful review~\cite{Maquety1998}.
Furthermore, recent reviews covering various aspects of the Coulomb GF can be found
in~\cite{DrakeBook}.
In Refs.~\cite{Papp2001,Papp2002,Shakeshaft2004,Feranchuk2011} the two-particle GF for the
Coulomb problem was derived through a convolution of two distinct one-electron Coulomb GFs.
These derived results were subsequently employed for the computation of Sturmian matrix elements.
It is noteworthy, however, that a comprehensive and detailed examination of the two-electron GF
was not carried out within the context of the aforementioned works.

In this paper, we extend the findings of Holster and Pratt~\cite{Hostler1963}
and Blinder~\cite{Blinder1981} to the case of two
non-relativistic electrons interacting
through Coulomb forces. This generalization is possible because on can separate
the motion of the electron pair into center-of-mass and relative motions.
However, several complexities arise. Firstly, the electrons repel each other,
preventing the formation of any bound states. Moreover, the electron pair can exist in either a singlet
state or one of three triplet states, which adds to the intricacy.
Additionally, the Pauli exclusion principle introduces extra limitations on the pair's wave function.

In contrast to the simpler Coulomb problem, the GF for the electron
pair depends on four variables instead of the usual two, making the calculations
more challenging. Most notably, the GF for the electron pair does
not neatly separate into center-of-mass and relative motion parts, which adds to the complexity.

In the subsequent sections of this paper, we outline our approach to overcoming
these challenges and obtaining the GF and the local density of states (LDOS) in terms
of quadratures of special functions, particularly Whittaker and Gamma functions.
This work offers a solution to these issues.

Our calculation of the GF for an electron pair and the LDOS
unfolds in three steps. In the first step, we neglect the electron's spin
and temporarily setting aside the Pauli exclusion principle. 
This phase involves the generalization
of the Coulomb GF to encompass a pair of particles governed
by the two-particle Schrodinger equation, influenced by the repulsive Coulomb potential.
As we progress to the second step, we take into account the presence of electron
spin and the presence of the Pauli exclusion principle.
This step involves deriving the even and odd components of the pair's GF,
preserving the requisite symmetries concerning the exchange of particles.
In the final step, we calculate the trace of the imaginary parts of both the even and odd
components of the pair's GF. This leads us to the calculation of the
local densities of states, corresponding to the singlet and triplet
states of an electron pair, respectively.

The paper is organized as follows.
In Section~\ref{Sec_NoSpin}, we calculate the GF while disregarding the presence of electron spins.
We derive the general expression for the two-particle GF, conduct a thorough analysis of
its properties, identify specific sets of arguments that lead to GF divergence,
and compute the local density of states in the limit of vanishing arguments.
Section~\ref{Sec_WithSpin} extends the results from the previous section to encompass the
scenario of two electrons, including their spins. This section also takes into account the
limitations imposed by the Pauli exclusion principle.
Section~\ref{Sec_Dys} is dedicated to the derivation of the Dyson equation for spin-independent potentials.
In Section~\ref{Sec_LDOS}, we calculate the LDOS as a function of inter-electron distance
and the pair's energy. In this section also analyze the pseudo-LDOS term, resulting from the
Pauli exclusion principle, which leads to the complete vanishing of the odd-part of the LDOS at~$r=0$.
Section~\ref{Sec_Dis} engages in discussions on several issues pertaining to the obtained results,
as well as potential possibilities for their experimental observation.
The paper is summarized in the Summary, followed by an appendix that offers
additional information and explanations to complement the main text.

\section{Green's function of two electrons in absence of spin effects \label{Sec_NoSpin}}

\subsection{General form of GF}

Consider two charged particles labeled 'a' and 'b,'
positioned at coordinates~$\bma$ and~$\bmb$, and both possessing a common mass~$m_e$ equivalent to the electron mass.
These particles carry charges of~$Z_a|e|$ and~$Z_b|e|$, where~$|e|$ denotes the elementary charge.
It's important to note that in this scenario, we are assuming spinless particles,
and as a consequence, the two-particle wave function
is not constrained by the Pauli exclusion principle. The Hamiltonian describing the system is
\begin{equation} \label{GF_Hp}
 \hH_p= -\frac{\hbar^2}{2m_e}\nabla^2_\bma -\frac{\hbar^2}{2m_e}\nabla^2_\bmb
 + \frac{Z_aZ_b}{4\pi\epsilon_0|\bma - \bmb|},
\end{equation}
where~$\epsilon_0$ represents the vacuum permittivity. Further in this
paper atomic units are introduced.
In this paper, our focus is on a pair of two electrons with charges~$(Z_a=Z_b=-1)$,
and they interact through the repulsive Coulomb interaction. For~$(Z_a=1, Z_b=-1)$,
the Hamiltonian~$\hH_p$ describes the positronium system.
Moreover, in the limit where~$m_a$ approaches infinity
while maintaining~$(Z_a=1, Z_b=-1)$, the Hamiltonian reduces to that of the hydrogen atom.

In center of mass~$\bmR=(\bma + \bmb)/2$ and relative~$\bmr=\bma - \bmb$
coordinates the Hamiltonian~$\hH_p$ separates
\begin{equation} \label{GF_HpCM}
 \hH_p = -c_K \nabla^2_\bmR + \left(-c_k\nabla^2_\bmr+ \frac{1}{r} \right),
\end{equation}
where in atomic units~$c_K=1/4$ and~$c_k=1$. Note that for the hydrogen atom~$c_k=1/2$.
The energy of the system sum of two terms:~$E_K = c_K K^2$ for the center-of-mass motion
and~$E_k$ of the relative motion. In absence of external potential there is~$E_k=c_kk^2$.

The Hamiltonian~$\hH_p$ does not possess any bound states, and its wave functions are delocalized. We have
\begin{equation} \label{GF_PhiCM}
 \Phi(\bmR,\bmr) = N_k e^{i\bmK\bmR} \psi_{klm}(\bmr).
\end{equation}
Here,~$N_k$ represents the normalization factor,~$\bmK$ is the wave vector for the center-of-mass motion,
and~$\psi_{klm}(\bmr)$ denotes the continuous states governed by the repulsive Coulomb Hamiltonian
\begin{equation} \label{GF_psi_klm}
 \psi_{klm}(\bmr) = R_{kl}(r)Y_{l}^{m}(\theta, \phi).
\end{equation}
In the above expression,~$Y_{l}^{m}(\theta, \phi)$ stands for the spherical harmonics in standard notation.
The parameters are defined as follows:~$k=\sqrt{E_k/c_k}$ represents the wave vector for relative
motion,~$l$ denotes the azimuthal quantum number,
and the radial function~$R_{kl}(r)$ is given by~\cite{LandauBook}
\begin{eqnarray} \label{GF_Phi_kl}
 R_{kl}(r) &=& \frac{C_{kl} (2kr)^l e^{ikr}}{(2l+1)!}\ _1F_1(i/k+l+1, 2l+2,-2ikr), \ \ \ \\
 \label{GF_Ckl}
 C_{kl} &=& 2ke^{-\pi/2k}|\Gamma(l+1+i/k)| \nonumber \\
        &=& \sqrt{\frac{8\pi k}{e^{2\pi/k}-1}}\prod_{s=1}^l \sqrt{s^2+\frac{1}{k^2}},
\end{eqnarray}
When~$l=0$, the product in Eq.~(\ref{GF_Ckl}) simplifies to unity.
The function~$_1F_1(\alpha, \gamma, z)$ is the confluent hypergeometric function
\begin{equation} \label{GF_F11}
 _1F_1(\alpha, \gamma, z) = 1 + \frac{\alpha}{\gamma} z + \frac{\alpha(\alpha+1)}{\gamma(\gamma+1)} z^2 + \ldots.
\end{equation}
For a fixed value of~$l$, the functions~$R_{kl}(r)$ are normalized according to the criterion~\cite{LandauBook}
\begin{equation} \label{GF_Norm}
 \int_0^{\infty} R_{kl}(r) R_{k'l}(r) r^2 dr = 2\pi \delta(k-k').
\end{equation}
In the~$(\bmR, \bmr)$ coordinates, the retarded (advanced) two-particle GF is defined as
\begin{eqnarray}
 g^{\pm}(\bmR_1,\bmR_2,\bmr_1,\bmr_2; E) = \nonumber \\ \label{GF_gpsi}
 \int \frac{d^3\bmK}{(2\pi)^3}\int_0^{\infty}\!\!\! dk
 \sum_{l=0}^{\infty}\sum_{m=-l}^l \frac{e^{i\bmK(\bmR_1-\bmR_2)}
 \psi_{klm}(\bmr_1) \psi_{klm}^*(\bmr_2) }
 {(E - c_K K^2) - c_k k^2 \pm i\eta} \ \ \\
 \label{GF_g1}
 = \frac{1}{(2\pi)^3}\int e^{i\bmK(\bmR_1-\bmR_2)}
 \tgc^{\pm}(\bmr_1,\bmr_2,\eK) d^3\bmK. \ \
\end{eqnarray}
Here, the term~$\eK$ is defined as:
\begin{equation} \label{GF_eK}
 \eK= E - c_K K^2.
\end{equation}
Additionally, we introduce a small positive parameter~$\eta > 0$. It is important to note that
the factor~$c_k$ is already accounted for in the Coulomb GF, as outlined in Eq.~(1.3) in Ref.~\cite{Hostler1964}.
The function~$\tgc(\bmr_1,\bmr_2;E)$ represents the one-particle GF for the Coulomb potential,
as discussed in~\cite{Hostler1963,Hostler1964}
\begin{eqnarray} \label{GF_gc}
 \tgc^+(\bmr_1, \bmr_2; E) = -\frac{\Gamma(1-i\nu)}{4\pi |\bmr_1-\bmr_2| } \times \nonumber \\
 \left(W^{1/2}_{i\nu}(U) \frac{\partial}{\partial V} \calM^{1/2}_{i\nu}(V)
 - \calM^{1/2}_{i\nu}(V) \frac{\partial}{\partial U} W^{1/2}_{i\nu}(U) \right),
\end{eqnarray}
where
\begin{eqnarray} \label{GF_UV}
 U &=& -ik \big(r_1 + r_2 + |\bmr_1-\bmr_2| \big), \\
 V &=& -ik \big(r_1 + r_2 - |\bmr_1-\bmr_2| \big),
\end{eqnarray}
$k = \sqrt{E/c_k}$ with~${\rm Im} \big\{ k \big \} > 0$ and~$\nu=-1/k$.
The functions~$\calM_{\kappa}^{\mu}(z)$ and~$W_{\kappa}^{\mu}(z)$ are the Whittaker functions
in notation used by Buchholtz~\cite{BuchholzBook,dlmfLink}. In Eq.~(\ref{GF_gc}),
the function~$\tgc$ represents the Coulomb GF that describes both attractive~($\nu > 0$)
and repulsive~($\nu < 0$) potentials~\cite{Blinder1981}. When~$\nu=0$, the function~$\tgc$ simplifies to the
GF of a free particle.

For specific applications, an alternative representation of the Coulomb GF proves to be more practical.
In this representation, the Coulomb GF is expanded using partial waves~\cite{DrakeBook}
\begin{equation} \label{GF_gc_pw}
 \tgc({\bm r}_1,{\bm r}_2;E) = \sum_{l=0}^{\infty}\sum_{m=-l}^{l} g_l(r_1,r_2;E) Y_{l}^{m}(\Omega_1)Y_{l}^{m*}(\Omega_2),
\end{equation}
where~$\Omega= (\theta,\phi)$ denotes angular variables, and the radial Coulomb GFs
are defined as~\cite{Hostler1964,Zon1969}
\begin{eqnarray} \label{GF_gl}
 g_l(r_1,r_2; E) &=& - \Gamma(1+l-i\nu)\frac{i\nu}{r_1r_2} \times \nonumber \\
 &\times & \calM^{l+1/2}_{i\nu}(\tm 2ikr_{\scriptscriptstyle <}) W^{l+1/2}_{i\nu}(\tm 2ikr_{\scriptscriptstyle >}),
\end{eqnarray}
where~$r_{\scriptscriptstyle <}=\min(r_1,r_2)$ and~$r_{\scriptscriptstyle >}=\max(r_1,r_2)$.
It is important to note that in the subsequent sections of this paper,
any form of~$\tgc({\bm r}_1,{\bm r}_2;E)$ is permissible.

The Coulomb GF presented in Eq.~(\ref{GF_gc}) is an analytic
function of energy in the complex energy plane. It exhibits a branch cut along the positive
real axis, which corresponds to the continuous spectrum.
In the case of negative energies, the Coulomb GF becomes a real function.

For an attractive Coulomb potential, the GF in Eq.~(\ref{GF_gc}) possesses simple poles at energies
corresponding to the singularities of~$\Gamma(1 - i\nu)$, specifically for~$1-i\nu = 0, -1, \ldots$.
This results in the discrete spectrum associated with the hydrogen atom.
However, for a repulsive potential, the Coulomb GF has no poles, and consequently, no bound states exist.
This same principle applies to the GF of an electron pair as described in Eq.~(\ref{GF_g1}),
where again, no poles are present, and therefore, no bound states are formed.

\subsection{Properties of two-particle GF}

When~$\bmR_1$ is distinct from~$\bmR_2$ and~$\bmr_1$ is different from~$\bmr_2$,
the integral over~$d^3\bmK$ in Eq.~(\ref{GF_g1}) exhibit no singularities.
The outcome of this integration depends on the signs of~$E$ and~$\eK$, leading to three distinct scenarios:
i) When both~$E$ and~$\eK$ are greater than zero, the Coulomb GF described in Eq.~(\ref{GF_g1})
has oscillatory behavior.
ii) In the case of~$E > 0$ and~$\eK < 0$, or iii) when both~$E$ and~$\eK$ are negative,
the Coulomb GF in Eq.~(\ref{GF_g1}) experiences exponential decay with increasing distance between the particles.
Specifically, for~$E > 0$
\begin{equation}
 g^{\pm}(\bmR_1,\bmR_2,\bmr_1,\bmr_2; E >0) = I^{\pm} + I^0,
\end{equation}
where
\begin{equation} \label{GF2_Ipm}
 I^{\pm} = \int_0^{\sqrt{E/c_K}} \frac{K\sin(KR_{12})}{2 \pi^2 R_{12}}
 \tgc^{\pm}(\bmr_1, \bmr_2,\eK) dK,
\end{equation}
and
\begin{equation} \label{GF2_I0}
 I^0 = \int_{\sqrt{E/c_K}}^{\infty}\frac{K\sin(KR_{12})}{2 \pi^2 R_{12}}
 \tgc^{0}(\bmr_1, \bmr_2,-|\eK|) dK,
\end{equation}
and~$R_{12}=|\bmR_1-\bmR_2|$.
In the case of~$E < 0$, the two-electron GF is a real valued function
expressed by a single term
\begin{eqnarray} \label{GF2_IEmin}
 g(\bmR_1,\bmR_2,\bmr_1,\bmr_2; E <0) = \nonumber \\
 \int_0^{\infty} \frac{K\sin(KR_{12})}{2 \pi^2 R_{12}} \tgc^{0}(\bmr_1, \bmr_2, \varepsilon_K) dK,
\end{eqnarray}
where
\begin{equation} \label{GF2_vK}
 \varepsilon_K = -|E| - c_K K^2 < 0,
\end{equation}
see Eq.~(\ref{GF_eK}).
In the equations above,~$\tgc^{\pm}$ represent the retarded and advanced
Coulomb GFs respectively,
while~$\tgc^{0}$ corresponds to the Coulomb GF associated with negative energies.
It's important to note that the integrals involving~$\tgc^{0}$ do not make any
contribution to the density of states.

Subsequently in this paper, we will adopt a simplified notation for the two-particle GF
\begin{equation}
 g^{\pm}(\bma_1, \bmb_1, \bma_2, \bmb_2;E) \equiv g(a_1b_1a_2b_2),
\end{equation}
and
\begin{eqnarray} \label{GF2_gab}
 g(a_1b_1a_2b_2) &=& \frac{1}{(2\pi)^3}\int e^{i\bmK(\bma_1+\bmb_1)/2}
 e^{-i\bmK(\bma_2+\bmb_2)/2}
 \times \nonumber \\
 && \tgc(\bma_1-\bmb_1, \bma_2-\bmb_2, \eK) d^3\bmK.
\end{eqnarray}
In this notation we do not make distinction between the retarder and advanced GFs.

\subsection{Divergences of~$g(a_1b_1a_2b_2)$}

In the majority of cases, the integral in Eq.~(\ref{GF2_gab}) converges.
However, for certain GF's arguments, it diverges,
leading to singular behavior of~$g(a_1b_1a_2b_2)$.
This integral diverges either for~$\bma_1+\bmb_1=\bma_2+\bmb_2$ or
for~$\bma_1-\bmb_1 = \bma_2-\bmb_2$. This occurs either
for vanishing exponent in Eq.~(\ref{GF2_gab}) or for equal arguments of Coulomb GF.
The cases having zero, one, or two nonzero values among the vectors~$\bma_1$,~$\bmb_1$,~$\bma_2$, 
and~$\bmb_2$ are detailed in Table~1. Instances with three nonzero arguments of GF are omitted.
\begin{table}
\begin{tabular}{|c|c|c|c|c|l|}
 \hline
 Group & $(\bma_1\bmb_1\bma_2\bmb_2)$ & $\bmr_1$ & $\bmr_2$ & $\bmR_1\tm \bmR_2$ & Integrand in Eq.~(\ref{GF_g1}) \\
 \hline
 0 & $(\bmZero\bmZero\bmZero\bmZero)$ & $\bmZero$ & $\bmZero$ & $\bmZero$ & $\tgc(\bmZero,\bmZero;\eK) d^3 \bmK $ \\
 \hline
 1 & $(\bmx\bmZero\bmx\bmZero)$ & $\bmx$ & $\bmx$ & $\bmZero$ & $\tgc(\bmx,\bmx;\eK) d^3 \bmK $ \\
 1 & $(\bmx\bmZero\bmZero\tm\bmx)$ & $\bmx$ & $\bmx$ & $2\bmx$ & $\tgc(\bmx,\bmx;\eK)e^{i\bmK\bmx} d^3 \bmK $ \\
 1 & $(\bmx\bmZero\bmZero\bmx)$ & $\bmx$ & $\tm\bmx$ & $\bmZero$ & $\tgc(\bmx,\tm\bmx;\eK) d^3 \bmK $ \\
 \hline
 2 & $(\bmx\bmy\bmx\bmy)$ & $\bmr$ & $\bmr$ & $\bmZero$ & $\tgc(\bmr,\bmr;\eK) d^3 \bmK $ \\
 2 & $(\bmx\bmy\bmy\tm\bmx)$ & $\bmr$ & $\bmr$ & $2\bmx$ & $\tgc(\bmr,\bmr;\eK)e^{i\bmK\bmx} d^3 \bmK $ \\
 2 & $(\bmx\bmy\bmy\bmx)$ & $\bmr$ & $\tm\bmr$ & $\bmZero$ & $\tgc(\bmr,\tm\bmr;\eK) d^3 \bmK $ \\
 \hline
\end{tabular}
\caption{Three groups of arguments of GF for which the integrals in Eqs.~(\ref{GF_g1}) and~(\ref{GF2_gab}) is diverge.
         Group~$0$ (all arguments vanish), group~$1$ (one non-zero argument~$\bmx$),
         and group~$2$ (two non-zero vectors,~$\bmx$ and~$\bmy$). We define~$\bmr = \bmx - \bmy$.
         Permutations causing simultaneous sign changes of~$\bmr_1$ and~$\bmr_2$,
         as well as~$\bmR_1$ and~$\bmR_2$, are excluded.}
\end{table}

For GF arguments provided in Table~1, it is necessary to use the spectral representation of the GF.
We illustrate this method to calculate~$g^{\pm}(0000;E) \equiv g^{\pm}_0(E)$.
From Eq.~(\ref{GF_gpsi}) we have
\begin{eqnarray}
 g_0^+(E) =\nonumber \\
 \label{GF2_g0}
 \int \frac{d^3\bmK}{(2\pi)^3} \int_{0}^{\infty}\!\!
 \sum_{l=0}^{\infty}\sum_{m=-l}^l \frac{|\psi_{klm}(\bmZero)|^2 dk}
 {E - c_K K^2 - c_k k^2 + i\eta}.
\end{eqnarray}
At~$\bmr=\bmZero$, all functions~$\psi_{klm}(\bmr)$ described in Eq.~(\ref{GF2_g0}) become zero,
except for those where~$l$ and~$m$ are both equal to zero. There is
\begin{equation} \label{GF2_fk}
 f(k) \equiv |\psi_{k00}(0)|^2 = \frac{8\pi k}{(2\pi)(4\pi)[\exp(2\pi/k)-1]},
\end{equation}
where~$k=\sqrt{E_k/c_k}$, see Eq.~(\ref{GF_HpCM}). The factor~$(4\pi)$
in the denominator of Eq.~(\ref{GF2_fk}) arises from the normalization of the spherical
function~$Y_{0}^{0}$, as described in Ref.~\cite{WikiY}.
Similarly, the factor of~$(2\pi)$ comes from the normalization of the radial functions
in Eq.~(\ref{GF_Norm}). Then we have
\begin{equation}\label{GF2_g01}
 g^{\pm}_0(E) = \int \frac{d^3\bmK }{(2\pi)^3}\int_0^{\infty}\!\! \frac{f(k)dk}{E - c_K K^2 - c_k k^2 \pm i\eta}.
\end{equation}
Using the the Dirac
identity:~$1/(x\pm i\eta) = {\cal P}(1/x) \mp i\pi\delta(x)$ to Eq.~(\ref{GF2_g01})
and integrating over angular variables we have
\begin{eqnarray} \label{GF2_g0a}
 {\rm Im} \big\{g_0^+(E) \big\} &=& \frac{(-\pi)}{2\pi^2}\int_0^{\infty}\!\! \int_0^{\infty}\!\!
 f(k) K^2 \times \nonumber \\ & \times & \delta(E- c_K K^2- c_k k^2) dK dk.
\end{eqnarray}
Next we introduce the polar coordinates~$K=\frac{t}{\sqrt{c_K}}\cos(\alpha)$,~$k=\frac{t}{\sqrt{c_k}}\sin(\alpha)$,
with~$0 \leq \alpha \leq \pi/2$ and the volume element:~$dK dk =t dt d\alpha/\sqrt{c_K c_k}$.
Using the identity
\begin{equation} \label{GF2_delta}
\delta(t^2-a^2) = \frac{1}{|2a|}\big [\delta(t-a) + \delta(t+a) \big],
\end{equation}
we obtain
\begin{eqnarray}
 {\rm Im} \big\{g_0^+(E>0) \big\} = \frac{(-\pi)}{2\pi^2 \sqrt{c_K^3 c_k}} \times \nonumber \\
 \int_0^{\pi/2}\!\! \int_0^{\infty}\!\!\!
 f[(t/\sqrt{c_k})\sin(\alpha)]t^3 \cos(\alpha)^2 \delta(E-t^2) dt d\alpha \nonumber \\
 \label{GF2_g0b}
 =\frac{(-\pi)E \Theta(E)}{4\pi^2 \sqrt{c_K^3 c_k}} \label{GF2_Img0}
 \int_0^{\pi/2} f[\sqrt{E/c_k}\sin(\alpha)]\cos(\alpha)^2 d\alpha, \ \ \ \ \
\end{eqnarray}
where~$\Theta(E)$ is the step function.
For negative energies in the second line of Eq.~(\ref{GF2_Img0})
there is~$\int_0^{\infty}\delta(-|E|-t^2) dt = 0$,
and~${\rm Im} \big\{g_0^+(E<0) \big\}$ vanishes.

The real part of~$g_0^+(E)$ is Hilbert transform of~${\rm Im} \big\{g_0^+(E) \big\}$,
and it diverges for all energies.
To circumvent this divergence, a cutoff energy~$W$ is introduced, beyond which the density of
states:~$\rho_0(E) = (-1/\pi){\rm Im} \big\{g_0^+(E) \big\}$ vanishes.
In physical terms, this signifies the finite width~$W$ of the energy band. Consequently, we have
\begin{equation} \label{GF2_Hilb}
 {\rm Re} \big\{ g_0^+(E) \big\} = -\frac{1}{\pi} {\cal P} \int_{0}^{W}
 \frac{{\rm Im} \big\{g_0^+(E') \big\}}{E-E'}.
\end{equation}
The above integral exists for finite~$W$ and it diverges for~$W \to \infty$.

\begin{figure} \includegraphics[width=8cm,height=8cm]{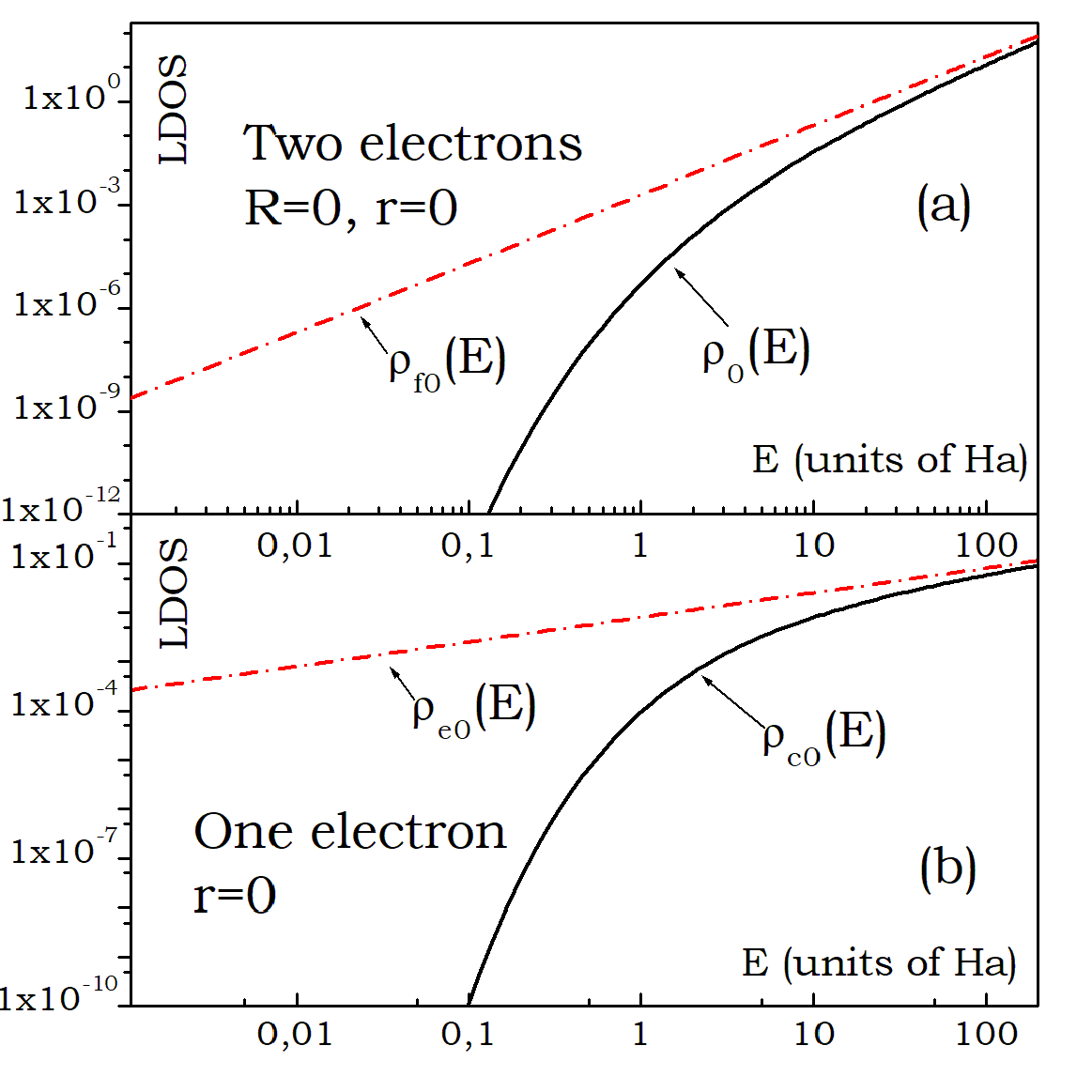}
 \caption{Panel (a), solid line:~$\rho_0(E)$
 for a pair of electrons in presence of Coulomb potential, as given in Eq.~(\ref{GF2_Img0}).
 Panel (a), dotted line:~$\rho_{f0}(E)$ for a pair
 of electrons in the absence of Coulomb interaction,
 as given in Eq.~(\ref{A_rho_f0b}).
 Panel (b), solid line:~$\rho_{c0}(E)$ for electron in a repulsive Coulomb
 potential, as given in Eq.~(\ref{A_rho_c0}). Panel (b),
 dotted line:~$\rho_{e0}(E)$ for free electron, see Eq.~(\ref{A_g1e}).
 In the presence of the Coulomb potential there is a non-zero overlap
 between pairs of electrons and a non-negligible probability of finding an
 electron at the center of the Coulomb potential.
 For two-electrons LDOS is in~$r_B^{-6}$ units, while for
 one-electron LDOS is in~$r_B^{-3}$ units.}
 \label{FigR0} \end{figure}

In Figure~\ref{FigR0}a, we present the local density of states denoted
as~$\rho_0(E) = (-1/\pi){\rm Im} \big\{g_0^+(E) \big\}$, as per Eq.~(\ref{GF2_g0b}),
represented by the solid line. The results are displayed on a log-log scale.
The primary observation from Figure~\ref{FigR0}a is that the LDOS for an electron pair subject
to Coulomb forces does not vanish at~$\bmR = \bmr = 0$. In other words, there is a non-zero
overlap between the two electrons, a manifestation of a purely quantum effect.
This behavior is intriguing, as classically, two electrons would repel each other and remain far apart.
This phenomenon, albeit somewhat mysterious, is also observable for a single electron
in a repulsive Coulomb potential.

In Figure~\ref{FigR0}b, we plot the local density of states~$\rho_{c0}(E)$ for a single electron
in a repulsive potential, as given in Eq.~(\ref{A_rho_c0}). 
As Figure~\ref{FigR0}b illustrates,~$\rho_{c0}(E)$
also does not vanish for any energy. This implies that there is a non-zero probability
that the electron overlaps with the center of the repulsive Coulomb potential,
a phenomenon rooted solely in quantum effects without a classical analogue.

The LDOS for an electron pair in Figure~\ref{FigR0}a and that for a single electron in a repulsive
potential in Figure~\ref{FigR0}b exhibit similar qualitative behavior.
For low energies, both LDOS profiles are negligibly small and gradually increase with energy.
At significantly high energy levels, a notable asymptotic behavior becomes evident.
The LDOS for the electron pair gradually converges towards the limiting local density of states~$\rho_{f0}(E)$,
which represents an electron pair in the absence of a Coulomb potential
and is defined in Eq.~(\ref{A_rho_f0b}).
This convergence is clearly illustrated in Figure~\ref{FigR0}a, where the dashed line closely
approaches the LDOS for the electron pair as energy increases.

Similarly, in this same high-energy limit, the density of states~$\rho_{e0}(E)$
for a single electron subjected
to a repulsive potential gradually approaches the density of states of a free electron~$\rho_{e0}(E)$,
as governed by Eq.~(\ref{A_g1e}). This convergence can be observed
in Figure~\ref{FigR0}b, where the dashed line closely aligns with the LDOS
for a single electron in the presence of a repulsive potential as energy levels rise.

\section{Green's Function for Two Electrons with Spin \label{Sec_WithSpin}}

In the preceding section, we conducted an analysis of the GF for a pair of
electrons interacting via Coulomb forces, while disregarding the influence of electron spins
and the constraints imposed by the Pauli exclusion principle. In this section, we extend our
study to encompass the GF of an electron pair, accounting for their spin
properties and the inherent limitations imposed by the Pauli principle.

Let's consider an arbitrary two-electron Hamiltonian, denoted as~$\hH$, which
is independent of electron spins, and let~$|\rm N\rangle$ represent one of its eigenstates.
We can then decompose~$|\rm N\rangle$ into two components
\begin{equation} \label{pa_N}
 |{\rm N}\rangle = |{\rm n}\rangle |\chi\rangle,
\end{equation}
Here,~${\rm n}$ encompasses all the quantum numbers that characterize state~$|{\rm n}\rangle$,
while~$|\chi\rangle$ signifies the state associated with the electron spins.

The wave function~$\langle \bma\bmb \sigma_a \sigma_b |{\rm N}\rangle$
separates on the spin-independent part~$\Psi_{\rm n}(\bma,\bmb)$ and
spin dependent function~$\chi(\sigma_a, \sigma_b)$, where~$\sigma_a, \sigma_b \in \{\uparrow, \downarrow\}$
are electrons spins. Because of the Pauli principle there is
\begin{equation} \label{pa_Psi}
 \Psi_{\rm n}(\bma,\bmb)\chi(\sigma_a, \sigma_b) =
 -\Psi_{\rm n}(\bmb,\bma) \chi( \sigma_b, \sigma_a).
\end{equation}
The aforementioned condition can be satisfied under two distinct scenarios.
Firstly, when the state function~$\chi(\sigma_a, \sigma_b)$ assumes the singlet state
form~$\frac{1}{\sqrt{2}}(|\uparrow\downarrow\rangle - |\downarrow\uparrow\rangle)$,
and the wave function~$\Psi_{\rm n}(\bma,\bmb)$ demonstrates even symmetry with respect to the exchange
of variables~$\bma$ and~$\bmb$. Alternatively, the condition is met when~$\chi(\sigma_a, \sigma_b)$
represents one of the triplet states:~$\frac{1}{\sqrt{2}}(|\uparrow\downarrow\rangle + |\downarrow\uparrow\rangle)$,
$|\uparrow\uparrow\rangle$, or~$|\downarrow\downarrow\rangle$,
and the wave function~$\Psi_{\rm n}(\bma,\bmb)$ exhibits odd symmetry with respect to~$\bma$ and~$\bmb$.

Let~$|\chi_s\rangle$ and~$|\chi_t\rangle$ be singlet and triplet states, respectively.
Then the Green's function~$\hG = (E-\hH)^{-1}$ is sum of two terms
\begin{widetext}\begin{equation} \label{pa_Geo}
 \hG^{\pm} = \sum_{\rm n}
 |\chi_s \rangle \langle \chi_s|\sum_{{\rm n}_e}
 \frac{|{\rm n}_e\rangle \langle {\rm n}_e|}{E - H \pm i\eta} +
 \sum_{\chi_t}|\chi_t \rangle \langle \chi_t|\sum_{{\rm n}_o}
 \frac{|{\rm n}_o\rangle \langle {\rm n}_o|}{E - H \pm i\eta}
 = \hG^{\pm}_e + \hG^{\pm}_o,
\end{equation} \end{widetext}
and the summation over~$\chi_t$ extends across all three triplet states.
In the position representation we have
\begin{eqnarray} \label{pa_Ge}
 G^{\pm}_e(a_1b_1a_2b_2) &=& \Lambda_s \sum_{{\rm n}_e}
 \frac{\Psi_{{\rm n}_e}(\bma_1,\bmb_1)\Psi^*_{{\rm n}_e}(\bma_2,\bmb_2)}{E-E_{{\rm n}_e} \pm i\eta}, \\
 \label{pa_Go}
 G^{\pm}_o(a_1b_1a_2b_2) &=& \Lambda_t \sum_{{\rm n}_o}
 \frac{\Psi_{{\rm n}_o}(\bma_1,\bmb_1)\Psi^*_{{\rm n}_o}(\bma_2,\bmb_2)}{E-E_{{\rm n}_o} \pm i\eta},
\end{eqnarray}
where~$\Lambda_s = \chi_s^{\dagger}\chi_s$,~$\Lambda_t =\sum_{\chi_t} \chi_t^{\dagger}\chi_t$.
By exploiting the symmetry properties of the functions~$\Psi_{{\rm n}_e}(\bma_1,\bmb_1)$
and~$\Psi_{{\rm n}_o}(\bma_1,\bmb_1)$
with respect to the exchange of coordinates~$\bma \leftrightarrow \bmb$ we find
\begin{eqnarray}
 \label{pa_Ge_ex}
 G_e(a_1b_1a_2b_2) = & G_e(b_1a_1a_2b_2) = & G_e(a_1b_1b_2a_2), \\
 \label{pa_Go_ex}
 G_o(a_1b_1a_2b_2) = & \tm G_o(b_1a_1a_2b_2) = & \tm G_o(a_1b_1b_2a_2).
\end{eqnarray}
As a consequence of Eq.~(\ref{pa_Go_ex}) there is
\begin{equation} \label{pa_Go_00}
 G_o(a_1b_100) = G_o(00a_2b_2) = G_o(0000) = 0,
\end{equation}
i.e., the the odd part of GF vanishes for~$\bmR=\bmr=\bmZero$.
In Eqs.~(\ref{pa_Ge}) and~(\ref{pa_Go}), the summation involves distinct functions,
namely,~$\Psi_{{\rm n}_e}(\bma,\bmb)$ for even states and~$\Psi_{{\rm n}_o}(\bma,\bmb)$ for odd states.
However, it is important to note that the denominators in both cases are identical
and equal to~$(E- c_K K^2 - c_k k^2 \pm i\eta)$.
The functions~$\Psi_{{\rm n}_e}(\bma,\bmb)$ and~$\Psi_{{\rm n}_o}(\bma,\bmb)$ can be readily
derived in the~$(\bmR,\bmr)$ coordinates,
as outlined in Eq.~(\ref{GF_PhiCM})
\begin{eqnarray}
 \label{pa_psi_e}
 \Psi_e(\bmR, \bmr) &=& \frac{1}{2} e^{i\bmK\bmR} \left[ \psi(\bmr) + \psi(-\bmr) \right], \\
 \label{pa_psi_o}
 \Psi_o(\bmR,\bmr) &=& \frac{1}{2} e^{i\bmK\bmR} \left[ \psi(\bmr) - \psi(-\bmr) \right].
\end{eqnarray}
where~$\psi(\bmr)\equiv N_k\psi(\bmr)_{klm}$ in Eq.~(\ref{GF_PhiCM}).
Inserting Eqs.~(\ref{pa_psi_e}) and~(\ref{pa_psi_o}) into Eqs.~(\ref{pa_Ge}) and~(\ref{pa_Go})
we find [see Eq.~(\ref{GF_g1})]
\begin{widetext}
\begin{eqnarray} \label{pa_geo}
 g^{\pm}_{e/o}(\bmR_1 \bmr_1 \bmR_2 \bmr_2) =
 \frac{1}{4} \frac{1}{(2\pi)^3} \left\{\begin{array}{c} \Lambda_s \\ \Lambda_t \end{array} \right \}
 \int d^3\bmK \sum_{l=0}^{\infty} \sum_{m=-l}^{l} \int_0^{\infty} dk
 \frac{e^{\bmK(\bmR_1-\bmR_2)}[\psi(\bmr_1) \pm \psi(\tm\bmr_1)][\psi^*(\bmr_2) \pm \psi^*(\tm\bmr_2)] }
 {E - c_K K^2 - c_k k^2 \pm i\eta} \nonumber \\
 =\frac{1}{4} \frac{1}{(2\pi)^3} \left\{\begin{array}{c} \Lambda_s \\ \Lambda_t \end{array} \right \}
 \int e^{\bmK(\bmR_1-\bmR_2)} \left[\tgc^{\pm}(\bmr_1,\bmr_2,\eK)
 \pm \tgc^{\pm}(\bmr_1,\tm\bmr_2,\eK)
 \pm \tgc^{\pm}(\tm\bmr_1,\bmr_2,\eK) + \tgc^{\pm}(\tm\bmr_1,\tm\bmr_2,\eK) \right] d^3\bmK \nonumber \\
 = \frac{1}{2} \left\{\begin{array}{c} \Lambda_s \\ \Lambda_t \end{array} \right \}
 \Big[ g^{\pm}(\bmR_1 \bmr_1 \bmR_2 \bmr_2,\eK) \pm g^{\pm}(\bmR_1 \bmr_1 \bmR_2 \tm\bmr_2,\eK) \Big].
\end{eqnarray} \end{widetext}
Here,~$\tgc^{\pm}(\bmr_1,\bmr_2)$ represents the Coulomb GF,
as defined in Eq.~(\ref{GF_gc}), and we have utilized the
property~$\tgc^{\pm}(\tm\bmr_1,\tm\bmr_2)= \tgc^{\pm}(\bmr_1,\bmr_2)$.
In this equation, the even function comprises the term~$\Lambda_s$,
whereas the odd function incorporates~$\Lambda_o$.

When the Coulomb GF is expanded in partial waves, see Eq.~(\ref{GF_gc_pw}),
the odd and even GFs in Eq.~(\ref{pa_geo}) can
be expressed in terms of spherical harmonic having even and odd angular quantum numbers only
\begin{widetext} \begin{eqnarray} \label{pa_pwe}
 g^{\pm}_{e}(\bmR_1 \bmr_1 \bmR_2 \bmr_2) &=&
 \frac{\Lambda_s}{(2\pi)^3} \sum_{l=0}^{\infty} \sum_{m=-l}^{l} Y_{2l}^{m}(\Omega_1)Y_{2l}^{m*}(\Omega_2)
 \int e^{\bmK(\bmR_1-\bmR_2)} g^{\pm}_{2l}(r_1,r_2;\eK) d^3 \bmK, \\
\label{pa_pwo}
 g^{\pm}_{o}(\bmR_1 \bmr_1 \bmR_2 \bmr_2) &=&
 \frac{\Lambda_t}{(2\pi)^3} \sum_{l=0}^{\infty} \sum_{m=-l}^{l} Y_{2l+1}^{m}(\Omega_1)Y_{2l+1}^{m*}(\Omega_2)
 \int e^{\bmK(\bmR_1-\bmR_2)} g^{\pm}_{2l+1}(r_1,r_2;\eK) d^3 \bmK.
\end{eqnarray} \end{widetext}

Returning to the~$(\bma, \bmb)$ coordinates, we obtain from Eq.~(\ref{pa_geo}) the following expressions
\begin{widetext}\begin{eqnarray}
 \label{pa_geab}
 g^{\pm}_e(a_1b_1a_2b_2;E) = \frac{\Lambda_{ss}}{2(2\pi)^3} \int e^{i\bmK(\bma_1 +\bmb_1 -\bma_2 -\bmb_2)/2}
 \big[ \tgc^{\pm}(\bma_1-\bmb_1,\bma_2-\bmb_2, \eK)
 + \tgc^{\pm}(\bma_1-\bmb_1,\bmb_2-\bma_2, \eK) \big] d^3\bmK, \\
 \label{pa_goab}
 g^{\pm}_o(a_1b_1a_2b_2;E) = \frac{\Lambda_{tt}}{2(2\pi)^3} \int e^{i\bmK(\bma_1 +\bmb_1 -\bma_2 -\bmb_2)/2}
 \big[ \tgc^{\pm}(\bma_1-\bmb_1,\bma_2-\bmb_2, \eK)
 - \tgc^{\pm}(\bma_1-\bmb_1,\bmb_2-\bma_2, \eK) \big] d^3\bmK.
\end{eqnarray}\end{widetext}
Note that when dealing with non-vanishing exponents and distinct arguments for the Coulomb GF,
the integration over the angular variables of the vector~$\bmK$ is straightforward,
as described in Eqs.~(\ref{GF2_Ipm})--(\ref{GF2_IEmin}).
However, in cases where the exponent's argument becomes zero or the arguments of the Coulomb GF are equal,
as summarized in Table~1, the integrals in Eqs.~(\ref{pa_geo})--(\ref{pa_goab})
diverge. In such instances, a different approach is required,
which is elaborated upon in Sections~\ref{Sec_NoSpin} and~\ref{Sec_LDOS}.

\section{Dyson equations for even and odd GFs \label{Sec_Dys}}

Let us examine the electron pair within the context of an external potential~$V(\bmr)$
that does not depend on spin and a two-electron interaction
represented as~$u(\bma,\bmb)$. The system's Hamiltonian is given by
\begin{equation} \label{Dy_HV}
 \hH = \hH_p + V(\bma) + V(\bmb) + u(\bma,\bmb) \equiv \hH_p + U(\bma,\bmb).
\end{equation}
Let~$\hG = \hG_e + \hG_o$ be the GF of the Hamiltonian in Eq.~(\ref{Dy_HV})
and~$\hg = \hg_e + \hg_o$ be the GF of the electron pair, see Eqs.~(\ref{pa_geab}) and~(\ref{pa_goab}).
Since~$\hG$ also separate into even and odd parts, see Eq.~(\ref{pa_Geo}),
we have~$\hG_e = \Lambda_s \widetilde{G}_e$,~$\hG_t = \Lambda_t \widetilde{G}_o$, and
$\hg_s = \Lambda_s \widetilde{g}_e$, and~$\hg_t = \Lambda_t \widetilde{g}_o$.
Then the Dyson equation~$\hG = \hg + \hg \hU \hG$ for the total GF reads
\begin{eqnarray}
 \big( \Lambda_s \widetilde{G}_e + \Lambda_t \widetilde{G}_o \big) = \nonumber \\
 \big( \Lambda_s \widetilde{g}_e + \Lambda_t \widetilde{g}_o \big) + \label{Dy_G0}
 \big( \Lambda_s \widetilde{g}_e + \Lambda_t \widetilde{g}_o \big) \hU
 \big( \Lambda_s \widetilde{G}_e + \Lambda_t \widetilde{G}_o \big).\ \ \
\end{eqnarray}
In the presence of spin-independent potentials, it holds that~$\Lambda_s\hU\Lambda_t=0$ due to the orthogonality
between singlet and triplet states. Consequently, Eq.~(\ref{Dy_G0}) can be split into two separate Dyson equations,
one for the odd GFs and another for the even GFs
\begin{eqnarray}
 \label{Dy_e} \hG_e &=& \hg_e + \hg_e \hU \hG_e, \\
 \label{Dy_o} \hG_o &=& \hg_o + \hg_o \hU \hG_o.
\end{eqnarray}
By taking the matrix element of both sides of equations~(\ref{Dy_e}) and~(\ref{Dy_o})
between the bra state~$\langle \bma_1\bmb_1 |$ and the ket state~$|\bma_2\bmb_2 \rangle$,
we obtain
\begin{eqnarray} \label{Dy_G1}
 G_{e/o}(a_1b_1a_2b_2) = g_{e/o}(a_1b_1a_2b_2) \nonumber \\
 + \iint g_{e/o}(a_1b_1a_3b_3) U(a_3b_3) G_{e/o}(a_3b_3a_2b_2) d^3a_3d^3b_3. \ \ \ \ \ \
\end{eqnarray}
When the operator~$\hU$ depends on electron spins, such as when it incorporates spin-orbit interactions,
it becomes necessary to apply the general formula as presented in Eq.~(\ref{Dy_G0}).

\section{LDOS for electrons pair \label{Sec_LDOS}}

For a system consisting of two electrons, the local density of states
can be derived from the GF as follows
\begin{equation}
 \varrho(\bma_1, \bmb_1;E) = (-1/\pi)\mathrm{Im}\ \mathrm{Tr}\{g(a_1b_1a_1b_1;E)\}.
\end{equation}
By taking limits~$\bma_2 \to \bma_1$,~$\bmb_2 \to \bmb_1$
in Eqs.~(\ref{pa_geab}) and~(\ref{pa_goab}),
setting~$\bmr = \bma_1 - \bma_2$ and using Eq.~(\ref{pa_Geo}) we have
\begin{equation} \label{LDOS_rho}
 \varrho(\bmr;E) = {\cal S}_s\varrho_e(\bmr;E) + {\cal S}_t\varrho_o(\bmr;E),
\end{equation}
where
\begin{widetext}\begin{eqnarray}
 \label{LDOS_rhoe}
 \varrho_e(\bmr;E) &=& -\frac{1}{2} {\rm Im} \Big\{
 \frac{1}{2\pi^3} \int_0^{K_m} \tgc^+(\bmr, \bmr; \eK)K^2 dK + \frac{1}{2\pi^3}\int_0^{K_m} \tgc^+(\bmr,-\bmr; \eK) K^2dK \Big\} \equiv
 \frac{1}{2} \big[ \varrho_{+}(\bmr;E) + \varrho_{-}(\bmr; E) \big], \\
 \label{LDOS_rhoo}
 \varrho_o(\bmr;E) &=& -\frac{1}{2}{\rm Im} \Big\{
 \frac{1}{2\pi^3} \int_0^{K_m} \tgc^+(\bmr, \bmr;\eK)K^2 dK - \frac{1}{2\pi^3}\int_0^{K_m} \tgc^+(\bmr,-\bmr; \eK) K^2dK \Big\} \equiv
 \frac{1}{2} \big[\varrho_{+}(\bmr;E) - \varrho_{-}(\bmr;E) \big],
\end{eqnarray} \end{widetext}
and~${\cal S}_s = \mathrm{Tr}{\Lambda_{ss}} = 1$,~${\cal S}_t =\mathrm{Tr}{\Lambda_{tt}} = 3$,
as detailed in Appendix.
In the equations above, we have employed the following definitions:~$K_m=\sqrt{E/c_K}$ and~$\eK=E-c_KK^2$,
as indicated in Eq.~(\ref{GF_eK}).
The limits of integration in Eqs.~(\ref{LDOS_rhoe}) and~(\ref{LDOS_rhoo})
arise from the condition that,
for~$\eK< 0$ the imaginary part of~$\tgc^+(\bmr_1, \bmr_2; \eK)$
vanishes, as discussed in Section~\ref{Sec_NoSpin} and Ref.~\cite{Hostler1964}. Finally we obtain
\begin{equation} \label{LDOS_rhot}
 \varrho(\bmr;E) = 2\varrho_{+}(\bmr;E) - \varrho_{-}(\bmr;E),
\end{equation}
For negative energies, both~$\varrho_{+}(\bmr; E)$ and~$\varrho_{-}(\bmr; E)$ become null,
as in such cases,~$\eK \leq 0$, and the imaginary component of the Coulomb GF vanishes for all values of~$K$.
However, for positive energies and finite values of~$r$, the situation is different.
The quantity~$\varrho_{-}(\bmr; E)$ can be determined through direct numerical integration~\cite{NRecBook}
of the Coulomb GF for~$\bmr_2 = -\bmr_1$
\begin{equation} \label{LDOS_rhom}
 \varrho_{-}(\bmr; E) = -\frac{1}{2\pi^3} \int_0^{K_m}
 \left[\frac{\Gamma(1-i\nu')}{8\pi r}W^{1/2}_{i\nu}(-4ik'r)\right] K^2 dK,
\end{equation}
as detailed in Eqs.~(\ref{GF_gc}) and~(\ref{GF_UV}). Here,~$k'=\sqrt{\eK/c_k}$,~$\nu_K = -1/k'$,
see Eqs.~(\ref{GF_eK}) and~(\ref{GF_gc}). In the above equation the Whittaker function
is well-defined for all~$r > 0$ and~$k > 0$.

To obtain~$\tgc^+(\bmr,\bmr; E>0)$ we compute~$\tgc^+(\bmr,\bmr_2; E>0)$ in the limit
as~$\bmr_2$ approaches~$\bmr$. Specifically, we express~$\bmr_2$ as~$\bmr + \bmdelta$,
where~$\bmdelta$ is a small vector oriented in an arbitrary direction.
It is important to note that we assume both~$\bmr$ and~$\bmdelta$ to be non-vanishing,
with the condition that~$\delta \ll r$. Then we have in Eq.~(\ref{GF_UV})
\begin{eqnarray}
 \label{LDOS_U}
 U &\simeq & -ik \big(r + r + \delta \cos(\varphi) + \delta \big) \equiv Z + A + \Delta, \\
 \label{LDOS_V}
 V &\simeq & -ik \big(r + r + \delta \cos(\varphi) - \delta \big) \equiv Z + A - \Delta,
\end{eqnarray}
where~$Z=-2ikr$,~$\varphi$ is the angle between~$\bmr$ and~$\bmdelta$,~$A=-ik\delta \cos(\varphi)$,
and~$\Delta = -ik \delta$. On applying the Taylor expansion
to~$W^{1/2}_{i\nu}(U) \equiv {\rm W}(U)$ and~$\calM^{1/2}_{i\nu}(V) \equiv {\rm M}(V)$ ($n=0$)
and their derivatives ($n=1$) we have
\begin{equation} \label{LDOS_TayW}
 \frac{d^n}{dZ^n}{\rm W}(Z + A + \Delta) \simeq \frac{d^n}{dZ^n}{\rm W}(Z) + (A + \Delta) \frac{d^{n+1}}{dZ^{n+1}} {\rm W}(Z),
\end{equation}
\begin{equation} \label{LDOS_TayM}
 \frac{d^n}{dZ^n}{\rm M}(Z + A - \Delta) \simeq \frac{d^n}{dZ^n}{\rm M}(Z) + (A - \Delta) \frac{d^{n+1}}{dZ^{n+1}} {\rm M}(Z).
\end{equation}
When substituting Eqs.~(\ref{LDOS_TayW}) and~(\ref{LDOS_TayM}) into Eq.~(\ref{GF_gc}) while retaining terms
at the lowest order in both~$A$ and~$\Delta$, we arrive at
\begin{widetext}\begin{eqnarray} \label{LDOS_gc1}
 \tgc^+(\bmr, \bmr + \bmdelta;E) \simeq -\frac{\Gamma(1-i\nu)}{4\pi\delta}
 \Big\{ \Big[{\rm W}(Z) \frac{d{\rm M}(Z)}{dZ} - {\rm M}(Z) \frac{d{\rm W}(Z)}{dZ} \Big] +
 A\Big[{\rm W}(Z)\frac{d^2{\rm M}(Z)}{dZ^2} -{\rm M}(Z)\frac{d^2{\rm {\rm W}}(Z)}{dZ^2}\Big] +\nonumber \\
 +\Delta \Big[{\rm M}(Z)\frac{d^2{\rm W}(Z)}{dZ^2} -2\frac{d{\rm M}(Z)}{dZ}\frac{d{\rm W}(Z)}{dZ}
 +{\rm W}(Z)\frac{d^2{\rm M}(Z)}{dZ^2} \Big] \Big\} + \ldots.
\end{eqnarray} \end{widetext}
The first bracket in Eq.~(\ref{LDOS_gc1}) represents the Wronskian of two Whittaker
functions~\cite{BuchholzBook,dlmfLink}
\begin{equation} \label{LDOS_Wro}
 {\cal W} \left\{W^{1/2}_{i\nu}(Z), \calM^{1/2}_{i\nu}(Z) \right\} = \frac{\Gamma(1)}{\Gamma(1-i\nu)}.
\end{equation}
The term linear in~$A$ vanishes since its is direct proportional to the derivative
of the Wronskian in Eq.~(\ref{LDOS_Wro}) in respect of~$Z$.
In the last bracket in Eq.~(\ref{LDOS_gc1}) the second order derivatives~$d^2{\rm W}(Z)/dZ^2$
and~$d^2{\rm M}(Z)/dZ^2$ are eliminated
by employing the Whittaker equation
\begin{equation} \label{LDOS_Witt}
\frac{d^2}{dZ^2} \left\{ \begin{array}{c} {\rm W}(Z) \\ {\rm M}(Z) \end{array} \right\} +
\left(-\frac{1}{4} + \frac{\kappa}{Z} \right)
\left\{ \begin{array}{c} {\rm W}(Z) \\ {\rm M}(Z) \end{array} \right\} =0,
\end{equation}
where~$\kappa=i\nu$. Since~$\Delta / \delta = -ik$ we obtain from Eq.~(\ref{LDOS_gc1})
\begin{widetext}\begin{equation} \label{LDOS_gc2}
 \tgc^+(\bmr, \bmr + \bmdelta;E) \simeq -\frac{1}{4\pi \delta} + \frac{ik\Gamma(1-i\nu)}{4\pi}
 \Big[ 2\frac{d{\rm M}(Z)}{dZ}\frac{d{\rm W}(Z)}{dZ} + 2{\rm M}(Z){\rm W}(Z)
 \Big(\frac{\kappa}{Z} - \frac{1}{4} \Big) \Big] + \ldots.
\end{equation}\end{widetext}
In the limit as~$\delta \to 0$, the divergence of~$\tgc^+(\bmr, \bmr + \bmdelta;E)$ arises primarily
from the first term in Eq.~(\ref{LDOS_gc2}), which is a real-valued quantity. For positive energies,
the second term in Eq.~(\ref{LDOS_gc2})
is a complex number with a non-vanishing imaginary component, thereby resulting in a finite LDOS in
Eqs.~(\ref{LDOS_rhoe}) and~(\ref{LDOS_rhoo}).
The integration over this term is accomplished through standard numerical techniques~\cite{NRecBook}.

\begin{figure} \includegraphics[width=8cm,height=10cm]{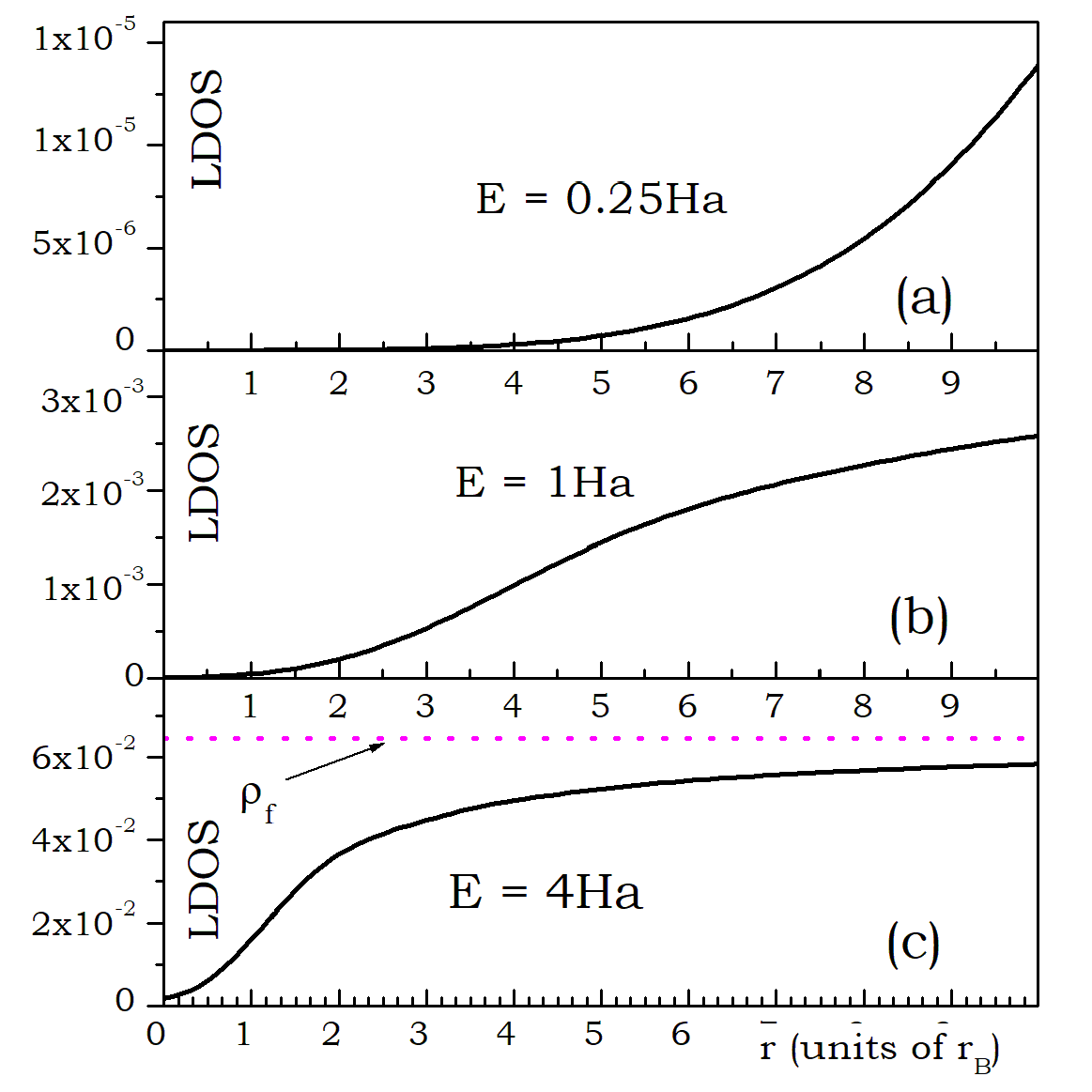}
 \caption{Local densities of states~$\varrho(\bmr;E)$,
 defined in Eq.~(\ref{LDOS_rhot})
 as a functions of inter-electron distance~$r$,
 calculated for three distinct energy values.
 In Panel (c), the dotted line represents~$\rho_{f}(E)$,
 as given in Eq.~(\ref{LDOS_rhof}), for~$E=4$ Ha.
 LDOS is in~$r_B^{-6}$ units.} \label{FigTot} \end{figure}

In Figure~\ref{FigTot}, we have plotted the local densities of states, as defined in Eq.~(\ref{LDOS_rhot})
for three different energy levels~$E$. In Figure~\ref{FigTot}a the energy is significantly
lower in comparison to the Hartree energy. The LDOS remains generally low within the considered range
of electron-electron distances~$r$ and virtually approaches zero in the region
of~$0 < r < 3$r$_B$, where~r$_B$ represents the Bohr radius.
Beyond this region, the LDOS exhibits an almost parabolic growth concerning distance,
following the relationship:~$\varrho(E) \propto (r - 3r_B)^2$.
Physically, this signifies that at lower energies, the electrons are considerably distant from each other.

For an energy of~$E=1$ Ha, as seen in Figure~\ref{FigTot}b, the LDOS displays a transition-like behavior.
It nearly diminishes for small electron-electron distances and then gradually increases for larger values of~$r$.
At approximately~$r=4.5$r$_B$, the LDOS curve shows an inflection point, suggesting the onset of LDOS saturation.
As energy increases, as shown in Figure~\ref{FigTot}c, we observe a logistic-like dependence of the LDOS on~$r$.
It nearly vanishes as~$r$ approaches zero and then rapidly increases, seemingly reaching saturation for~$r > 6$r$_B$.
The behavior of the LDOS for large values of~$r$ indicates that saturation occurs for LDOS values
corresponding to those of a free electron pair. According to Eq.~(\ref{A_rho_f0b}) we have
\begin{equation} \label{LDOS_rhof}
 \rho_f(E) \approx 2 \times \frac{E^2}{16 \pi^3}.
\end{equation}
The factor of~$2$ arises from the summation over two spin directions, as we assume a common spin for both
electrons in the case of spinless electrons.

\begin{figure} \includegraphics[width=8cm,height=8cm]{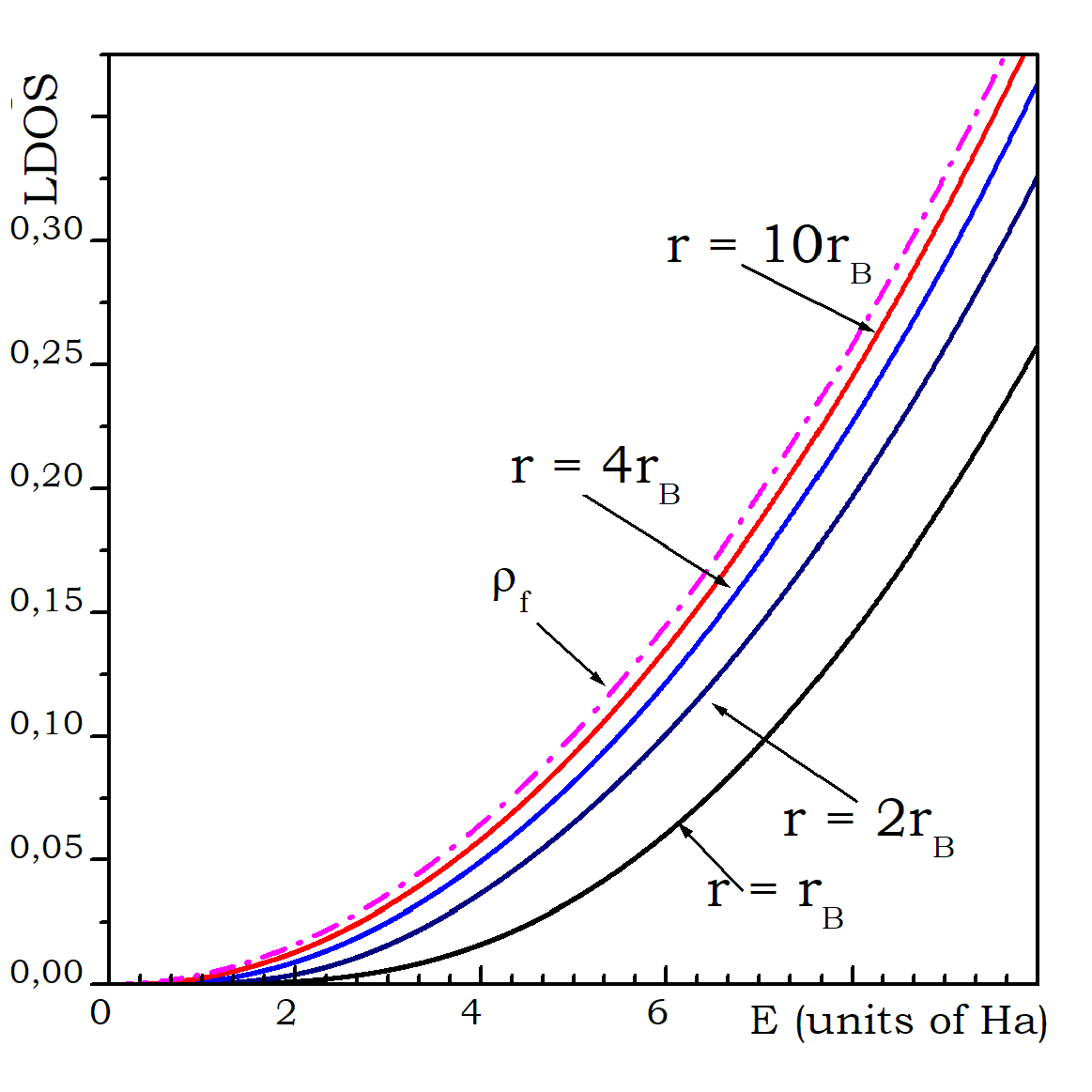}
 \caption{Solid lines: Local densities of states~$\varrho(\bmr;E)$
 defined in Eq.~(\ref{LDOS_rhot})
 as a functions of energy,
 calculated for four values of inter-electron distance~$r$. Dotted line:~$\rho_{f}(E)$,
 as given in Eq.~(\ref{LDOS_rhof}), as a function of energy.
 LDOS is in~$r_B^{-6}$ units.
 Note that for large values of~$r$, the local density of states~$\varrho(\bmr;E)$
 approaches~$\rho_{f}(E)$.} \label{FigEr} \end{figure}

To confirm the saturation level in Figure~\ref{FigTot}c, we calculate in Figure~\ref{FigEr} the LDOS at
four values of the electron-electron distance~$r$ as a function of energy, represented by solid lines,
and compare them with~$\rho_{f}(E)$ in Eq.~(\ref{LDOS_rhof}), denoted by a dotted line.
As seen in Figure~\ref{FigEr}, for high energies and large electron-electron distances,
the LDOS of the electron pair does not depend on~$r$ and tends to the uniform
electron density of a pair of free electrons as given in Eq.~(\ref{LDOS_rhof}).

In conclusion, based on the results presented in Figures~\ref{FigTot} and~\ref{FigEr},
we observe the following quantitative
behavior of the local density of states as a function of energy and electron distance. 
For low energies and small~$r$,
the LDOS is negligibly small but still nonzero. By increasing energy or electron distance,
the LDOS increases following a power-law trend up to an inflection point.
The position of this point depends on energy and shifts to lower values of~$r$ as~$E$ increases.
Ultimately, for large energies or distances, the LDOS is uniform and corresponds to the LDOS of
a pair of electrons in the absence of the Coulomb potential.

The physical picture corresponding to the above dependence of LDOS on energy and inter-electron
distance is as follows.
For every energy, there is an area of low electron density around~$r=0$,
where the Coulomb repulsion between electrons is strong, and the electrons do not overlap.
The size of this area decreases with energy. By increasing~$r$, the Coulomb repulsion weakens,
and electrons start to overlap, leading to an increase in LDOS. Finally, at large distances,
the Coulomb repulsion between electrons vanishes, and they behave as free particles.

\begin{figure} \includegraphics[width=8cm,height=8cm]{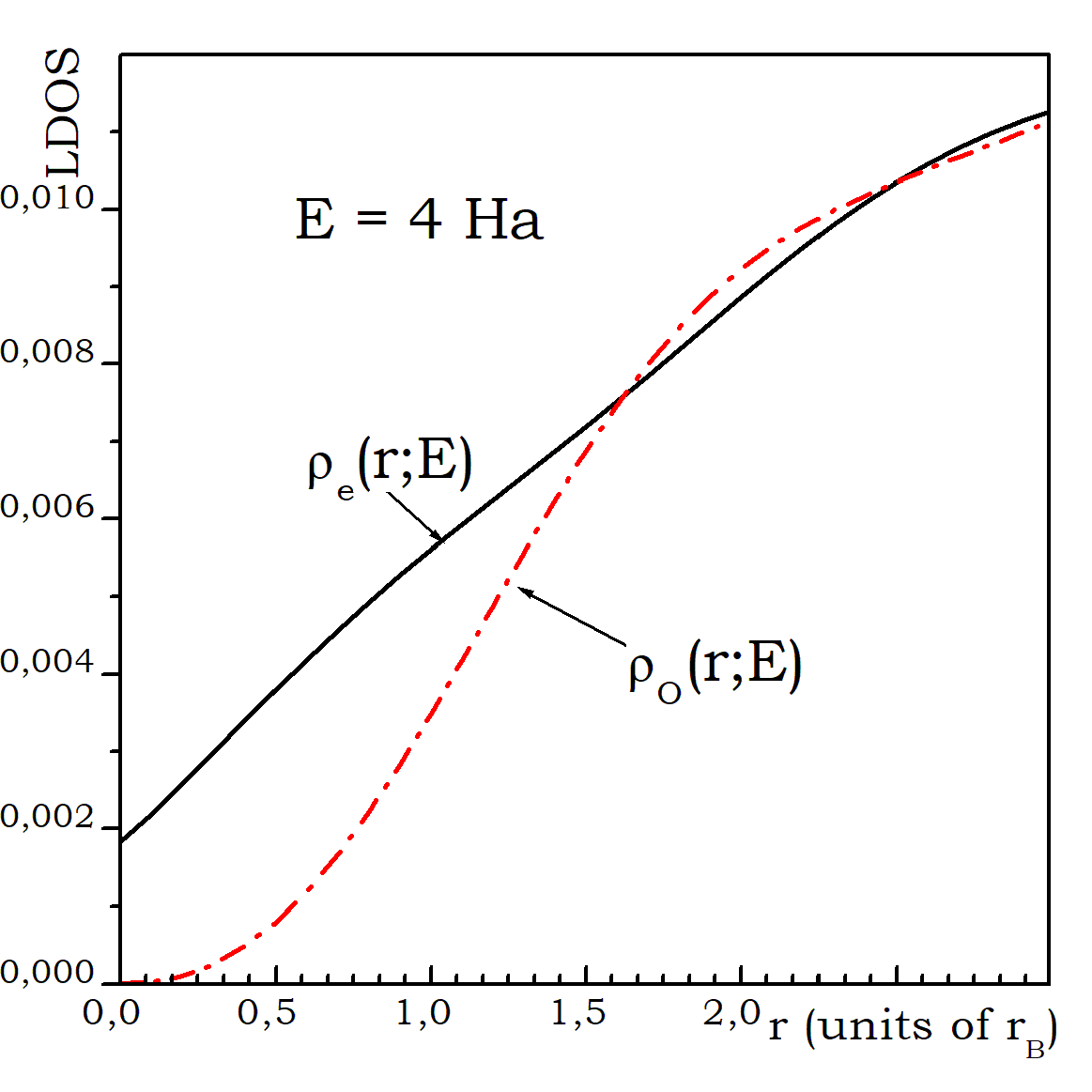}
 \caption{Even part~$\varrho_e(\bmr;E)$ and odd part~$\varrho_o(\bmr;E)$ of local density of states
 defined in Eqs.~(\ref{LDOS_rhoe}) and~(\ref{LDOS_rhoo}), respectively, as functions of the inter-electron
 distance~$r$. Both quantities are in~$r_B^{-6}$ units. Note that~$\varrho_o(\bmr;E)$
 precisely vanishes at~$r=0$ due to the Pauli exclusion principle,
 while~$\varrho_e(\bmr;E)$ remains finite in this limit.
 For large values of~$r$, both quantities are in close proximity to each other.}
\label{FigEO} \end{figure}

It is important to observe that the LDOS shown in Figures~\ref{FigTot} and~\ref{FigEr}
does not approach zero as~$r$ tends to~$0$. However, for low energies, it is exceptionally
small and can be considered negligible.

Let us now examine the even and odd components of the local density of states,
as defined in Eqs.~(\ref{LDOS_rhoe}) and~(\ref{LDOS_rhoo}),
corresponding to LDOS for singlet and triplet states, respectively.
In Figure~\ref{FigEO}, the local densities of states~$\varrho_e(r;E)$
and~$\varrho_o(r;E)$ are plotted for~$E=4$ Ha,
focusing on small inter-electron distances.
The primary distinction between them lies in their behavior as~$r$ approaches~$0$.
In this limit, the odd component of the LDOS precisely converges to zero,
as implied by Eqs.~(\ref{pa_Go_00}) and~(\ref{LDOS_rhoo}).
Conversely, the even component of LDOS remains finite in this limit.
As~$r$ increases, a significant disparity between~$\varrho_e(r;E)$ and~$\varrho_o(r;E)$
persists until~$r > 2.5$r$_B$, beyond which the two quantities become nearly identical.

\begin{figure} \includegraphics[width=8cm,height=8cm]{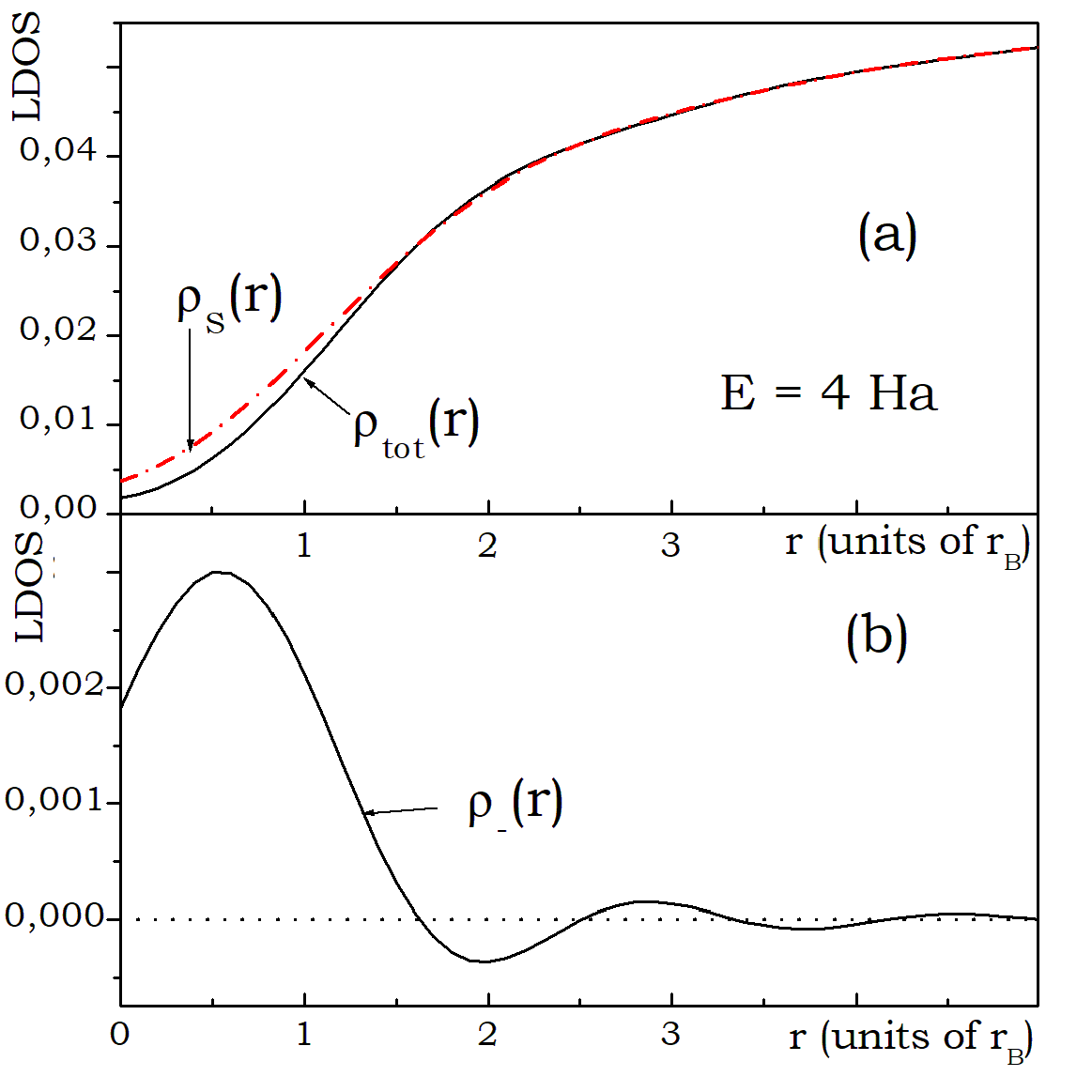}
 \caption{Panel (a): Local density of states~$\varrho_e(\bmr;E)$
 given in Eq.~(\ref{LDOS_rhot}) (solid line)
 and local density of states in the absence of spin effects~$\varrho_{s}(r;E)$ given in Eq.~(\ref{LDOS_rhos}),
 as functions of the inter-electron distance~$r$. The difference between these quantities,
 denoted as~$\varrho_{-}(r;E)$ and defined in Eq.~(\ref{LDOS_rhom}), is plotted in Panel (b).
 Note the different scale in Panel (b). The term~$\varrho_{-}(r;E)$ arises from the
 Pauli principle, ensuring the vanishing of the odd part of LDOS at~$r=0$
 as discussed in the text. LDOS is in~$r_B^{-6}$ units.} \label{FigSch} \end{figure}

The discrepancy between the local densities of states~$\varrho_e(r;E)$
and~$\varrho_o(r;E)$ in Figure~\ref{FigEO}
arises as a consequence of many-body effects,
specifically, it is linked to the term~$\varrho_{-}(r;E)$ in
Eqs.~(\ref{LDOS_rhoe})--(\ref{LDOS_rhom}).
For a more in-depth investigation of this phenomenon, Figure~\ref{FigSch} facilitates a comparison
between~$\varrho(r,E)$ and the LDOS in the absence of spin effects, which is defined as
\begin{equation} \label{LDOS_rhos}
 \varrho_{s}(r;E) = 2 \varrho_{+}(r;E),
\end{equation}
The factor of~$2$ in Eq.~(\ref{LDOS_rhos}) appears due to summation over two spins, see Eq.~(\ref{LDOS_rhof}).
As shown in Figure~\ref{FigSch}a, for~$E=4$ Ha, the deviation between~$\varrho(r;E)$ and~$\varrho_{s}(r;E)$
is primarily localized at small values of inter-electron separation~$r$.
In Figure~\ref{FigSch}b, the quantity~$\varrho_{-}(r;E)$, predominantly concentrates at small~$r$
distances and exhibits a decaying and oscillatory behavior,
diminishing beyond~$r > 3r_B$. It is worth noting that~$\varrho_{-}(r;E)$ cannot be considered a genuine
density of states as it lacks positive definiteness.

This quantity should be interpreted as a many-body contribution to the overall
density of states,~$\varrho(r;E)$, arising from electron correlations. Its emergence can be attributed to the Pauli
exclusion principle, which imposes specific spatial symmetries on electron wave functions concerning
the exchange of particles. The presence of~$\varrho_{-}(r;E)$ ensures that, for triplet states,
both the wave function and the LDOS precisely vanish at~$\bmR = \bmr = \bmZero$,
indicating that two electrons in any triplet state cannot occupy the same spatial point.

It is noteworthy that the Pauli exclusion principle demonstrates a considerable
potency in comparison to the Coulombic interaction between electrons.
The latter allows for a minor yet non-negligible degree of electron overlap at the point
where the positions of the electrons~$\bmR = \bmr = \bmZero$ coincide,
as visually depicted in Figure~\ref{FigR0}. 
The necessity for the term~$\varrho_{-}(r;E)$ to exist is a direct consequence of the
presence of a non-zero density of states at~$r=0$. In the event that~$\varrho_{+}(r;E)$ were to vanish
at~$r=0$ in Eq.~(\ref{LDOS_rhoo}), the Pauli exclusion principle would be
intrinsically satisfied by the~$\varrho_{+}(r;E)$ term alone,
rendering any corrections at~$r=0$ unnecessary.
However, owing to the finite value of~$\varrho_{+}(r;E)$ for~$r \to 0$, the supplementary
term~$\varrho_{-}(r;E)$ becomes indispensable to ensure the complete suppression
of electron overlap at~$r=0$.

The Pauli exclusion principle has a limited range of influence, as shown in Figure~\ref{FigSch}b.
The term~$\varrho_{-}(r;E)$, which reflects the Pauli exclusion effect, is predominantly
concentrated at small inter-electron distances~$r$ and diminishes rapidly beyond a certain point.
Consequently, for large values of~$r$, the density of states~$\varrho_t(r;E)$
converges to~$\varrho_s(r;E),$ as depicted in Figure~\ref{FigSch}a.
Similarly, at larger~$r,$ there's no distinction between~$\varrho_{+}(r;E)$
and~$\varrho_{-}(r;E)$, as demonstrated in Figure~\ref{FigEO}.

\section{Discussion \label{Sec_Dis}}

In the previous sections, various issues related to the obtained results have already been discussed.
In this section, we shift our focus to other aspects of the considered problem.

In our method, we analytically evaluate the double summation over~$l$ and~$m$
and integrate over~$k$ in Eq.~(\ref{GF_gpsi}) using the Coulomb GF.
This yields the GF expressed as an integral over~$\tgc(\bmr_1, \bmr_2; \eK)$
with energy~$\eK$ depending on the wave vector~$\bmK$.
Alternatively, it's possible to reverse the order of integration by considering the variable~$\bmK$ first.
This alternative approach leads to the same GF results
as those presented in Eq.~(\ref{GF_g1}), albeit in a more tedious way.

The results in this paper pertain to three-dimensional space but can be extended to one-dimensional
systems as the closed-form Coulomb GF is known for~$1D$~\cite{Meixner1933}.
Generalization is also possible for
systems with dimensions~$D=5,7,\ldots$ due to a discovered relationship between Coulomb~GF
in different dimensions~\cite{Hostler1970}
\begin{equation} \label{Di_tgcD}
 \tgc(x,y;E;D+2) = -\frac{1}{2\pi y} \frac{\partial}{\partial y} \tgc(x,y;E;D).
\end{equation}
This relation applies for integer values of~$D \geq 1$. However, there is no closed-form Coulomb~GF
for~$2D$ systems. Thus, our method extends to systems with dimensions~$D=2,4,6,\ldots$ only when
employing the partial-wave expansion of the Coulomb GF,
as shown in Eq~.(\ref{GF_gc_pw}).
The two-dimensional equivalent of Eq.~(\ref{GF_gc_pw}) can be found in Ref.~\cite{Hostler1970}.

The LDOS calculations in Figures \ref{FigTot}--\ref{FigSch} used an analytical expression for the Coulomb GF,
as shown in Eq.~(\ref{GF_gc}). However, alternative representations of the Coulomb GF are also applicable.
These include the partial waves expansion as indicated in Eq.~(\ref{GF_gc_pw}),
the momentum representation developed by Schwinger~\cite{Schwinger1964},
or representations involving hyperbolic functions,
as referenced in Ref.~\cite{Maquety1998}. It is worth noting that deriving an explicit closed-form expression
for the GF for two particles seems unattainable, and in any case,
one typically arrives at the GF in the form of a definite integral.

Regarding the numerical aspects of the calculation, the Whittaker functions in Eq.~(\ref{GF_gc})
are computed using series expansions for small arguments~\cite{dlmfLink}, and for large arguments,
they are obtained from corresponding asymptotic expansions, as described in the reference~\cite{dlmfLink}.

Additionally, the guidelines provided in Ref.~\cite{Pearson2017} for calculating
confluent hypergeometric functions have been taken into consideration.
Accuracy was ensured by monitoring the Wronskian of the two Whittaker functions
in Eq.~(\ref{LDOS_Wro}) for each parameter~$k$,~$\nu$, and~$r$ in Eq.~(\ref{GF_gc}).
For fixed~$\nu$ this Wronskian value should remain constant,
independent of the arguments of the Whittaker functions,
providing a dependable metric for assessing accuracy.

The method presented in this paper is applicable, with some modifications,
to two-particle molecules subject to attractive Coulomb potentials
such as excitons, positronium, and muonium, among others.
The key distinctions in the GF calculations for these systems,
compared to the results in this paper, are as follows:
i) Existence of bound states due to attractive Coulomb interactions in these systems.
ii) Absence of the Pauli exclusion principle, because of the opposite charge signs in these systems.
iii) Transition from a negative parameter~$\nu$ in the Coulomb GF in Eq.~(\ref{GF_gc}) for repulsive Coulomb
potentials to a positive parameter~$\nu$ for attractive potentials, leading to different
behavior of the Whittaker functions within the Coulomb GF.
In addition to these primary differences, other system-specific factors may come into play,
such as varying electron and hole effective masses in excitons or finite recombination
lifetimes in positronium. Nevertheless, in principle, our approach remains applicable to these systems as well.

The method outlined in our paper is specifically designed for two-particle systems.
It excels at separating the motion of two electrons into center-of-mass and relative motion
(as detailed in Eq.~(\ref{GF_HpCM})), making it universally applicable to
all two-particle systems with interactions depending on~$|\bmr_1 - \bmr_2|$.
However, this approach cannot
be extended to systems with three or more particles.
The fundamental reason is that the motion separation, which forms the basis of our approach, 
is infeasible in systems involving three or more particles. Therefore, our method is not well-suited 
for computing the GF in multi-particle systems.

When either~$\bmR_1 = \bmR_2$ or~$\bmr_1 = \bmr_2$, we calculate the real part of the GF
as a Hilbert transform of the imaginary part with a cutoff energy~$W$, see Eq.~(\ref{GF2_Hilb}).
In the case of a pair of non-relativistic electrons in a vacuum,
there isn't a natural or intrinsic cutoff energy. One potential choice for a cutoff is~$W \approx m_ec^2 = 0.511$ MeV,
which exceeds the characteristic energy of a pair, typically on the order of~$27.21$ eV, by four orders of magnitude.
For electrons confined in a quantum dot with barrier height~$U_0$,
choosing~$W \approx U_0$ is a reasonable cutoff. However, the selection of a cutoff energy is model-specific
and should be rigorously justified based on the system's characteristics.
Cutoff determination lacks a universal method; it hinges on the unique attributes of the model under investigation.

For experimental observation of LDOS in Figures~\ref{FigTot}--\ref{FigSch},
one viable approach is measuring the LDOS of an electron pair within quantum dots.
Figure~\ref{FigTot} demonstrates that the dot's size should exceed the effective
Bohr radius~$r_B^*$
of electrons within the dot, given by~\cite{WikiQDot}
\begin{equation} \label{Di_rb}
 r_B^* = r_B \kappa \frac{m_e}{\mu^*},
\end{equation}
where~$\kappa$ represents the material's dielectric constant, and~$\mu^*$ is the reduces effective mass of
electron pair.
Typically,~$r_B^*$ falls in the range of approximately~$50-100$\AA.
In such a system, the presence of dot barriers may minimally affect pair motion.
The LDOS can be observed using techniques like scanning tunneling microscopy, scanning tunneling potentiometry,
scanning tunneling spectroscopy, or tip-enhanced Raman spectroscopy, see e.g.~\cite{Schintke2004,Kroger2008}.
In addition to the total LDOS, it is feasible to measure the LDOS for singlet and triplet states
of electrons within the quantum dot. To control the spin states of an electron pair,
various methods can be employed, such as electron tunneling between electrodes,
optoelectronic and magneto-optical techniques, or injecting electrons in specific spin states from an external source.

Regardless of the experimental method used, the system allows for the measurement of the following effects:
i) The overall shape of the LDOS concerning energy and inter-electron distance, as shown in Figure~\ref{FigTot}.
ii) The free-particle limit of the LDOS at sufficiently high energies, demonstrated in Figures~\ref{FigTot} and~\ref{FigEr}.
iii) The distinction between the LDOS for singlet and triplet states of the electron pair, as seen in Figure~\ref{FigEO}.
iv) A complete absence of the LDOS at the origin, i.e., when the inter-electron distance is~$r=0$.
These effects remain robust even in the presence of confinement resulting from the quantum dot structure.

\section{Summary \label{Sec_Sum}}

In this paper, we extended the Coulomb Green's function to a system involving two electrons
interacting through repulsive Coulomb forces. We derived closed-form expressions for the GF,
which are represented by one-dimensional integrals, as shown in Eqs.~(\ref{GF2_Ipm}) to~(\ref{GF2_IEmin}).
It's important to emphasize that these equations are valid for systems where electron spins are not considered.
The obtained GF has no poles, and no bound states exist.

For positive energies, the obtained GF comprises a complex oscillatory term
with a non-vanishing imaginary component and a real term that decays exponentially with
inter-electron distance. For negative energies, the GF is real
and also decays exponentially with inter-electron distance.

For certain combinations of GF arguments, specifically when~$\bmR_1= \bmR_2$ or~$\bmr_1=\bmr_2$,
the integrals for the real parts of the GF diverge. However,
the imaginary parts of the GF remain finite,
resulting in a finite local density of states. We examined these cases in detail,
including situations where~$\bmR_1 = \bmR_2 = \bmr_1 = \bmr_2 = \bmZero$, as illustrated in Figure~\ref{FigR0}.
It has been discovered that for any pair energy, the LDOS at this point remains finite,
indicating a non-zero overlap of electrons wave functions. This phenomenon lacks a classical counterpart.
We also considered the scenario where~$\bmr_1 = \bmr_2 \neq \bmZero$, as described in Eq.~(\ref{LDOS_gc2}).

In the subsequent phase, we further generalize the results to include the spins of electrons and account
for the influence of the Pauli exclusion principle. In this context, we discovered that
the GF is composed of two terms,
each characterized as either odd or even concerning the exchange of particles, as described in Eq.~(\ref{pa_Geo}).
We derived closed-form expressions for the even and odd GF as sums and differences of GFs
with the appropriate arguments, as outlined in Eqs.~(\ref{pa_geo}) to~(\ref{pa_goab}).

We also discovered that for spin-independent potentials, the Dyson equation separates
into two distinct equations, one for even and the other for odd GFs, respectively.
After obtaining the GF for an electron pair in the presence of spins, we proceeded
to calculate the LDOS for the system.
The LDOS is a sum of contributions from the even and odd parts of the GF,
as described in Eq.~(\ref{LDOS_rho}). In Figures~\ref{FigTot} to~\ref{FigSch},
we computed the LDOS as a function of inter-electron distance and the pair's energy.
Additionally, we separately calculated the odd and even contributions to the LDOS,
highlighting the significance of the Pauli exclusion principle.

We further investigated the pseudo-local density of states,
denoted as~$\varrho_{-}(\bmr;E)$, which signifies the many-body contribution to the GF
and guarantees the total suppression of the local density of states at~$r=0$.
The necessity of including this term arises from the non-zero local density of
states at~$r=0$, as depicted in Eq.~(\ref{LDOS_rhoo}) and Figure~\ref{FigEO}.
This term exhibits a relatively limited spatial extent and diminishes
as the inter-electron distances increase.

We hope our paper facilitates a deeper understanding of the non-relativistic electrons
interacting through repulsive Coulomb forces and underscores the significance of the Pauli
exclusion principle in few-electron systems.

\appendix
\section{}

The retarded GF for a free particle with positive energies is
\begin{equation} \label{A_g1e}
 g^+_{e}(\bmr_1,\bmr_2;E) = -\frac{1}{4\pi r} \exp(i\sqrt{E}r),
\end{equation}
where~$r=|\bmr_1 -\bmr_2|$. In the limit~$r \to 0$ there is~$\rho_{e0}(E) = \sqrt{E}/(4\pi^2)$.
For the Coulomb GF corresponding to a repulsive potential, we obtain, see Eq.~(\ref{GF2_g01}),
\begin{equation} \label{A_rho_c0}
 \rho_{c0}(E) = -\frac{1}{\pi}\int_0^{\infty} \frac{f(k) dk}{E - c_kk^2 + i\eta}
 = \frac{1}{2\pi}\frac{f(\sqrt{E})}{\sqrt{E}},
\end{equation}
where~$c_k=1$ and we have utilized Eq.~(\ref{GF2_delta}).

We will now proceed to calculate the LDOS for a system of two electrons in absence of
Coulomb interaction.
At the specific point~$\bmR=\bmr =\bmZero$, the LDOS for a noninteracting electron pair,
denoted as~$\rho_{f0}(E)$, is given by
\begin{equation}
\label{A_rho_f0a}
 \rho_{f0}(E) = \frac{-1}{\pi}\frac{1}{4\pi^4}\int_0^{\infty}\int_0^{\infty}
 \frac{k_a^2 k_b^2 dk_a dk_b}{E - \frac{1}{2} k_a^2 -\frac{1}{2} k_b^2 + i\eta}.
\end{equation}
By introducing polar coordinates, where~$(k_a, k_b) \to (t, \alpha)$ and applying Eq.~(\ref{GF2_delta}), we obtain
\begin{eqnarray}
\rho_{f0}(E) &=& \frac{1}{2\pi^4}
\int_0^{\pi/2}\int_0^{\infty} t^5\cos^2(\alpha)\sin^2(\alpha)^2 \delta(2E-t^2) dt \nonumber \\
&=& \label{A_rho_f0b}
\frac{E^2}{16\pi^3} \Theta(E),
\end{eqnarray}
where the step function~$\Theta(E)$ ensures that the integral is non-zero only 
for~$E > 0$, and vanishes for~$E < 0$.

For arbitrary positions~$\bmR$ and~$\bmr$, the LDOS for a pair of non-interacting electrons
also can be obtained in a closed form. The retarded GF is
\begin{equation} \label{A_gf2_fa}
 g^+_f(\bmR,\bmr;E) = \frac{1}{(2\pi)^6}
 \int \frac{e^{i\bmK \bmR} e^{i{\mathbf k} \bmr} d^3 \bmK d^3 {\mathbf k}}
 {E - c_K K^2 - c_k k^2 + i\eta}.
\end{equation}
By integrating over~$d^3 {\mathbf k}$, we arrive at the expression
\begin{equation} \label{A_gf2_fb}
 g^+_f(\bmR,\bmr;E) = \frac{1}{(2\pi)^3}\int e^{i\bmK \bmR} g^+_{1e}(\bmr, \eK) d^3 \bmK,
\end{equation}
where~$\eK= E - c_K K^2$, and~$g^+_{1e}(\bmr, E)$ is given in Eq.~(\ref{A_g1e}).
For~$E < 0$, the imaginary part of~$g^+_0(\bmr;E)$ vanishes, leading to a reduction of the LDOS for negative energies.
For~$E>0$ we have
\begin{eqnarray} \label{A_gf2_fc}
 {\rm Im} \big \{g^+_f(\bmR,\bmr;E>0) \big\} = \nonumber \\
 = -\frac{1}{2\pi^2}
 \int_0^{\infty} \left[ \frac{\sin(r\eK)}{4\pi r} \right] \left(\frac{K \sin(KR)}{R}\right) dK.
\end{eqnarray}
To evaluate the above integral, we use the identity~(2.5.25.1) in Ref.~\cite{PrudnikovBook}
\begin{eqnarray} \label{A_2_5_25_1}
 \int_0^a \sin(r\sqrt{a^2-K^2}) \cos(qK) dK = \nonumber \\
 = \frac{\pi}{2}\frac{ar}{\sqrt{q^2+r^2}} J_1(a\sqrt{q^2+r^2}),
\end{eqnarray}
By differentiating both sides of Eq.~(\ref{A_2_5_25_1}) with respect to~$dq$ 
and introducing~$E' = E/c_K = 4E$, we derive the following from Eq.~(\ref{A_gf2_fc}).
\begin{eqnarray} \label{A_rho_fb}
 \varrho_f(\bmR,\bmr;E) &=& \frac{1}{16\pi^3} \left[-\frac{E' J_0\big (t\sqrt{E'}\big)}{2 t^2}
 +\frac{E' J_2\big (t\sqrt{E'} \big)}{2 t^2} \right. + \nonumber \\
 && \left. +\frac{\sqrt{E'}J_1\big(t\sqrt{E'} \big)}{t^{3}} \right],
\end{eqnarray}
where~$t= \sqrt{R^2 + r^2}$, and~$E > 0$.
It is worth noting that the limits~$\bmR \to \bmZero$ and~$\bmr \to \bmZero$
yield the LDOS as given in Eq.~(\ref{A_rho_f0b}).

To calculate the traces over~$\Lambda_{ss}$ and~$\Lambda_{tt}$ in Eqs.~(\ref{LDOS_rhoe})
and~(\ref{LDOS_rhoo}), we consider a singlet state~$|s \rangle$ for an electron pair
and a vector of triplet states~${\mathbf t} =[|t_1 \rangle, |t_2 \rangle, |t_3\rangle]$ for the pair.
In this context, the trace of~$\Lambda_s =|s \rangle \langle s|$ equals unity.
Meanwhile, the matrix~$\Lambda_t = {\mathbf t} \cdot {\mathbf t}^{\dagger}$ is a~$3 \times 3$
identity matrix, and its trace equals three.

\end{document}